\documentclass[fleqn,10pt]{wlscirep}

\usepackage{color}

\title{Percolation-based precursors of transitions in extended systems}

\author[1]{V\'{\i}ctor Rodr\'{\i}guez-M\'{e}ndez}
\author[1]{V\'{\i}ctor M. Egu\'{\i}luz}
\author[1]{Emilio Hern\'{a}ndez-Garc\'{\i}a}
\author[1,*]{Jos\'e J.\ Ramasco}
\affil[1]{Instituto de F\'{\i}sica
Interdisciplinar y Sistemas Complejos IFISC (CSIC-UIB),
Campus Universitat de les Illes Balears, E-07122 Palma de
Mallorca, Spain.}

\affil[*]{jramasco@ifisc.uib-csic.es}

\begin{abstract}
Abrupt transitions are ubiquitous in the dynamics of complex
systems. Finding precursors, i.e. early indicators of their
arrival, is fundamental in many areas of science ranging from
electrical engineering to climate. However, obtaining warnings
of an approaching transition well in advance remains an elusive
task. Here we show that a functional network, constructed from
spatial correlations of the system's time series, experiences a
percolation transition way before the actual system reaches a
bifurcation point due to the collective phenomena leading to
the global change. Concepts from percolation theory are then
used to introduce early warning precursors that anticipate the
system's tipping point. We illustrate the generality and
versatility of our percolation-based framework with model
systems experiencing different types of bifurcations and with
Sea Surface Temperature time series associated to El Ni\~{n}o
phenomenon.
\end{abstract}

\begin{document}

\flushbottom
\maketitle

\thispagestyle{empty}

\section*{Introduction}

The occurrence of sharp transitions to different states or
regimes during the evolution of complex systems is a phenomenon
of major importance both from the fundamental point of view and
for practical implementations of control and management.
Examples of such abrupt changes can be found in
ecology\cite{Scheffer2001}, economy \cite{Yan2010}, electrical
engineering \cite{Dobson2007}, physiology
\cite{VandeLeemput2014} or climate
\cite{Dakos2008,Thompson2011,Lenton2012}.
Detecting with sufficient anticipation the approach to a
critical or tipping point has thus become an important issue.
Early-warning signals have been introduced and tested in recent
works \cite{Scheffer2012,Wang2012b,Dakos2015}, including
experimental verification in living and environmental systems
\cite{Veraart2012e,Wang2012b,Quail2015}. These methods rely on
the loss of resilience occurring generically when dynamical
systems approach most (although not all) types of bifurcation
points \cite{Scheffer2009,Thompson2011,Dakos2015}. Recovery
rates from perturbation become small, leading to {\sl critical
slowing down} of the dynamics, increased memory, long temporal
autocorrelations, and to growth of the temporal variance
\cite{Scheffer2009,Thompson2011,Dakos2015}. From a dynamical
viewpoint, these critical slowing down phenomena appear when
the eigenvalue of the Jacobian matrix describing the rate of
relaxation towards the attractor approaches zero close to
bifurcation points. In many cases, particularly when different
spatial parts of the system are coupled by diffusion-like
processes, the increase in temporal correlation is accompanied
by the growth of spatial correlations
\cite{Guttal2009,Dakos2010,Dakos2011}. This, with the
associated increase in spatial variance and response functions,
is actually a standard method to characterize phase transitions
in thermodynamic physical systems \cite{Chaikin1995}.

The consideration of spatial correlations has led to a novel
perspective for finding transition precursors through the use
of correlation or functional networks
\cite{Tsonis2004,Eguiluz2005,Timme2014,Donges2009}. These
networks are built by identifying spatial units as nodes in a
graph, measuring the correlation among all pairs of them, and
keeping the most significant ones as link weights. Several
network-based precursors have been proposed. Specifically the
values of the degree (number of connections per node),
assortativity (degree-degree correlations), clustering (average
density of triangles) and kurtosis raise when
approaching a tipping point
\cite{VanderMheen2013d,Tirabassi2014,Feng2014}.

Here we show an important additional property of these
functional networks: As internal correlations increase,
networks evolve from a low to a high connectivity state and,
\emph{before} reaching its maximum link density at the
bifurcation point, a percolation-like transition occurs in the
network topology. Concepts from graph percolation theory
\cite{Stauffer1994,Newman2010} can thus be imported to
characterize this transition. Importantly, metrics can be
defined that act as early warnings for this percolation
transition, which itself is a precursor of the dynamic
transition. The validity of these general ideas is tested
by analyzing model systems displaying different types of
bifurcations: steady and oscillatory, continuous and
discontinuous. In all cases, we observe the occurrence of
percolation transitions before the dynamical one, and we
characterize it with quantities that can be used as
early-warning signals. Our approach uses only time series from the elements of an extended system, without the need of specific knowledge
about the underlying dynamics. Thus, it is suited to analyze
observational data for which little or no modeling insight is
available. This property of our framework is illustrated
by applying it to temperature data from the Pacific ocean associated
to El Ni\~{n}o phenomenon.

\section*{Methods}

\subsection*{Functional networks and precursors}

The time evolution of extended dynamical systems is
described by time-dependent spatial fields. Let $\psi({\mathbf x},t)$ be
one of such fields, and consider a suitable discretization of
it  $\{\psi({\mathbf x}_l,t_k)\}_{l,k}$, defining a time-series at
discrete times $t_k$, $k=1,...,R$, from each spatial location
${\mathbf x}_l$, $l=1,...,N$. The construction of functional networks implies to
compute the Pearson correlation from the time series at every pair of locations:
\begin{equation}
\rho_{a b} =\frac{\sum_k p_a(t_k) p_b(t_k)}
   {\sqrt{\left(\sum_k p_a(t_k)^2\right)\left(\sum_k p_b(t_k)^2\right)}} \ ,
\label{pearson}
\end{equation}
where $p_l(t_k)\equiv \psi({\mathbf x}_l,t_k) - \frac{1}{R}\sum_k
\psi({\mathbf x}_l,t_k)$ is the deviation of the field from its
temporal mean at each location. A network in which nodes are
the spatial locations ${\mathbf x}_l$ is defined by assigning links
between pairs of nodes $({\mathbf x}_a,{\mathbf x}_b)$ for which the Pearson
correlation in equation (\ref{pearson}) is higher than a predefined
threshold $\gamma$: $\rho_{ab}>\gamma$.

To study percolation in these functional networks, as the
control parameter $p$ approaches transition points $p_d$, we
measure several metrics: a) $S_1$, the relative size of
the largest connected component, i.e. the fraction of nodes
that are in the largest cluster. It abruptly changes from a
value close to 0 to a value close to 1 at the percolation
point. b) The average size of the clusters excluding the
largest one. This quantity is maximal at the percolation point
\cite{Stauffer1994}. It can be calculated as
\begin{equation}
\sum_{s} s \, c_s =
\frac{1}{N}\sum_{s} s^2 \, n_s \equiv \langle s^2 \rangle,
\end{equation}
where
the sum runs over all cluster sizes $s$ excluding the largest one,
and $c_s$ is the fraction of nodes belonging to clusters of
size $s$, $c_s = s \, n_s/N$ ($n_s$ is the number of clusters of
size $s$ present in the system \cite{Newman2010}). $c_s$ gives also the
probability that a randomly chosen node pertains to a cluster
of size $s$.

The standard percolation indicators, $S_1$ and $\langle s^2
\rangle$ are not, however, the best precursors. We show below
that better anticipation can be obtained by exploiting the interplay between the probabilities $c_s$ and the coming transition. In random graphs it is possible to perform analytic calculations on the behavior of $c_s$. As new links are  added at random, the percolation occurs when the mean degree $\langle k \rangle$, which acts as the control parameter $p$, equals one. Before this, the probabilities $c_s$ that a randomly chosen node
belongs to a component of size $s$ can be written as
\begin{equation}
c_{s} = {e^{-s\langle k\rangle}(s\langle
k\rangle)^{s-1}}/{s!} .
\end{equation}
These probabilities have a maximum when the
mean degree, which in this case is the parameter $p$, is $\langle k\rangle=p_s = \frac{s-1}{s}$. The
succession $\{p_s\}$ of location of maxima of $c_s$ converges
to the percolation point $p_\infty=1$ for increasing component
size $s$, but for low $s$ these maxima could be quite far from
the percolation point, and in fact they anticipate it. We show
below that the early-warning character of the maxima of $c_s$
holds also true in non-random functional networks obtained from systems undergoing very different dynamic transitions. This is due to the generic increase in the dynamical correlations of the elementary units of the system that the approach of a global tipping point brings. The sequence of peaks in $c_s$ offers thus a general and versatile tool to predict potential changes in the dynamics at a global scale.

\subsection*{Percolation and transitions in model systems}

We analyze here three different extended dynamical systems
displaying different types of bifurcations. The first two
examples experience steady bifurcations (a discontinuous
saddle-node and a continuous pitchfork). The third case, the
Lorenz'96 system, experiences a variety of transitions being
the first one an oscillatory Hopf bifurcation between a steady
state and traveling waves. Further transitions occur when
changing the control parameter leading to low-coherence
spatio-temporal chaos. In all cases we add random noise to the
deterministic model. This represents the unavoidable stochastic
fluctuations to which real systems are always subjected, and
provide the necessary statistics to have well-defined spatial
correlation functions. In all cases percolation occurs in the
associated functional networks, providing robust early-warning
signals of the approaching transition.

\section*{Results}

\subsection*{A lake eutrophication model}


\begin{figure}
\begin{center}
\includegraphics[width=0.7\columnwidth]{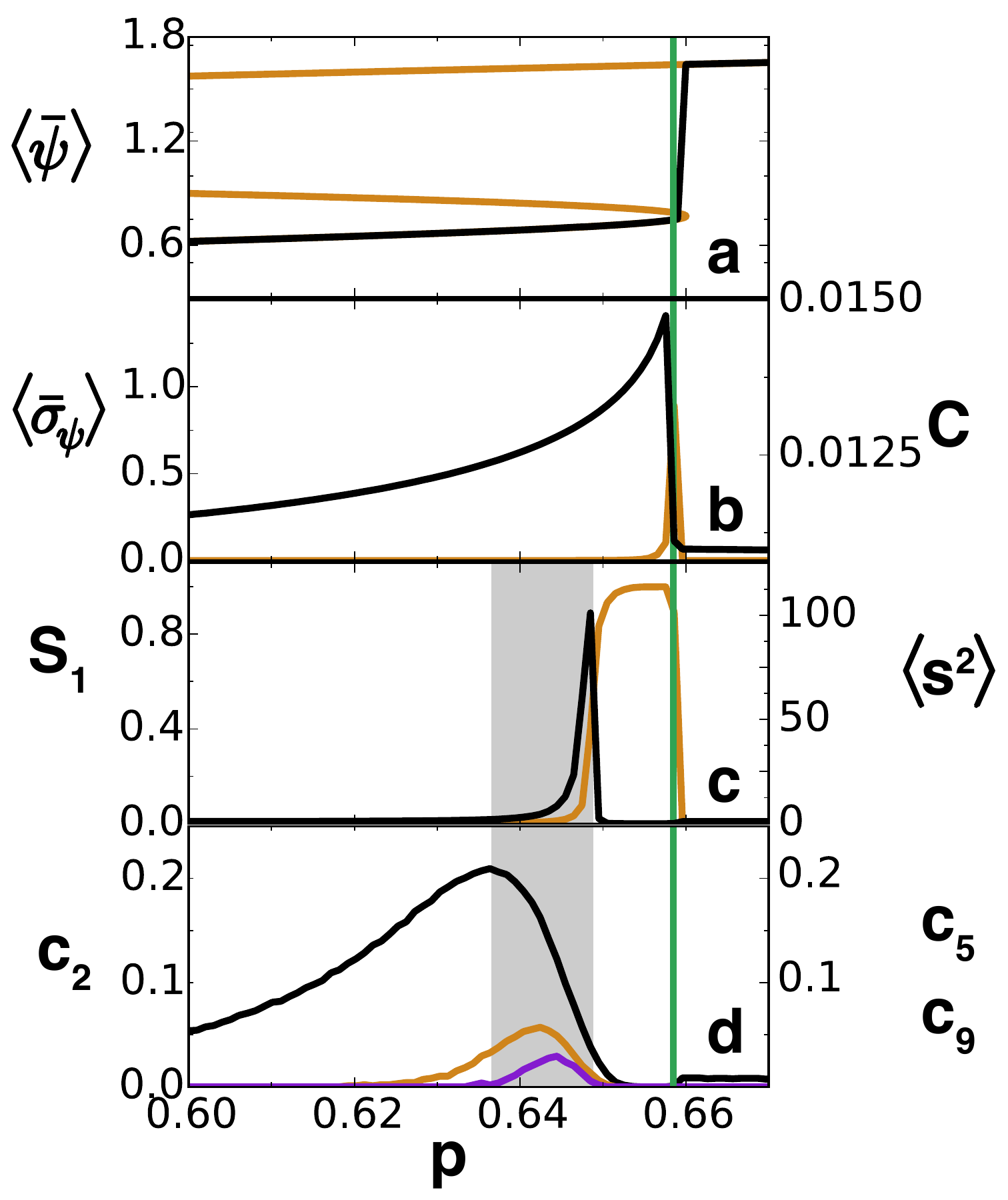}
\caption{Transition precursors for the LE model, equation (\ref{Lk}).
In a, the steady
homogeneous phosphorus concentration $\psi$ (orange) and the numerically
obtained spatial average (black, further averaged over $R=1000$ temporal snapshots for each value of $p$)
as a function of the control parameter $p$ which
is slowly increased from low to high values. In b,
$\langle\bar{\sigma_\psi}\rangle$, the spatial standard deviation (black) of $\psi$ used in Refs. \cite{Scheffer2009,Guttal2009,Dakos2010} as a transition precursor is displayed
averaged over $R=1000$ temporal snapshots.
In orange, the clustering of the functional network built with threshold $\gamma = 0.21$  also used as a precursor in Refs. \cite{VanderMheen2013d,Tirabassi2014}.
In c, the relative size of the giant component, $S_1$ (orange),
and the average size of the leftover clusters ($\langle s^2 \rangle$, black).
d) The probabilities $c_2$ (black), $c_5$ (orange), and $c_9$ (purple) are shown.
In all the panels, the vertical green line marks the position of the observed abrupt transition
($p_d=0.658$). The grey area indicates the anticipation in
parameter space gained over previous methods by using the peak of $c_2$ as precursor of the percolation transition. All curves
have been further averaged over 100 realizations of the random noise and initial
conditions.} \label{figLE}
\end{center}
\end{figure}

As a first example, we consider a lake eutrophication (LE)
model which is a spatial version of a description of
phosphorous recycling in a lake \cite{Carpenter1999}. It
suffers a paradigmatic abrupt transition associated to a
saddle-node (SN) bifurcation, namely a transition between two
contrasted states for the phosphorus concentration
$\psi({\mathbf x},t)$ in the lake: low concentrations leading to clear
water, and excess of phosphates leading to turbid water. The
state of the system is given by the two-dimensional field
$\psi({\mathbf x},t)$, representing the amount of phosphorus in the
lake, evolving according to
\begin{equation}
\frac{\partial\psi({\mathbf x},t)}{\partial t} = p - b\psi({\mathbf x},t) +
r f(\psi) + \epsilon\nabla^2 \psi({\mathbf x},t) + \eta({\mathbf x},t) \ .
\label{Lk}
\end{equation}
$f(\psi)=\psi^8/(\psi^8+1)$ is a nonlinear response of the lake
sediments to phosphorus, $p$ is the nutrient input rate, taken
here as the control parameter. $\epsilon$ is the strength of
diffusive spatial coupling, and $\eta$ represents an additive
stochastic perturbation uncorrelated in space and time. We take
$b=r=1$ and $\epsilon=1.2$. Space is discretized as a square
lattice of $N=70\times70=4900$ grid points separated by $dx=1$.
These will be the nodes of the functional network. The
Laplacian is discretized with the simplest finite differences
scheme and the deterministic terms in equation (\ref{Lk}) are
integrated with a 4th order Runge-Kutta method of time step
$dt=0.05$ after which the $\eta$ term is implemented by adding
an independent random number uniform in $[-a,a]$ to each
lattice site (we use $a=0.125$). We approach from the left the
SN bifurcation occurring at $p=p_{SN}=0.660$, above which the
clear-water low-phosphorus state existing for $p<p_{SN}$
ceases to exist and the lake jumps to an eutrophicated
high-phosphorus state. The stochastic perturbation makes the
jump to occur at a value of $p$, $p_d$, slightly below the SN
value. On average (see Fig. \ref{figLE}), we find $p_d\approx
0.658$. This value is sufficiently close to $p_{SN}$ as to
display the enhancement of correlations and slowing down which
are at the basis of our method and of other early-warning
methodologies. In this paper we show only results obtained when
slowly increasing the control parameter $p$. When decreasing
$p$ from higher values hysteresis occurs and a different SN
bifurcation is encountered at a lower $p'_d\approx 0.389$. The
sequence of precursors encountered when approaching this lower
transition point is similar to the one shown here.

The increase of the spatial variance was used in
\cite{Guttal2009,Dakos2010} as transition
precursor. This spatial variance is defined for the discretized
field $\psi({\mathbf x}_l,t_k)$ in the asymptotic statistically steady
state as:
\begin{equation}
\bar{\sigma_{\psi}} = \sqrt{\frac{1}{N}\sum\limits_{l=1}^{N}(\psi({\mathbf x_{l}},t) - \bar\psi(t))^2} \ ,
\label{signmaSpatial}
\end{equation}
where $\bar\psi(t) =
\frac{1}{N}\sum\limits_{l=1}^{N}\psi({\mathbf x_{l}},t)$ is the
average over the spatial nodes. When the system approaches
$p_d$, the raise of spacial variance can be observed for the LE
model (Fig. \ref{figLE}b).

As explained above, we constructed functional networks by
assigning links between locations among which spatial
correlations (as measured by Pearson correlation) are larger
than a threshold for which we use $\gamma=0.21$. We take averages over
$R=1000$ temporal snapshots.
The spatial correlations,
computed from equation (\ref{pearson}), increase and lead to a
growth of the link density in the vicinity of the critical
point as indicated by the precursors proposed in
\cite{VanderMheen2013d,Tirabassi2014,Feng2014}. In fact one of
these network precursors, the clustering coefficient, is
plotted in Fig. \ref{figLE}b (orange) and has a peak at the
dynamical transition. But we will show that there is also a
percolation transition, and to capture it we studied how the
size of the giant component, $S_1$, and the average size of the
leftover clusters $\langle s^2 \rangle$ change with $p$ (see
Fig. \ref{figLE}c). Note that $\langle s^2 \rangle$ has a peak
at $p=0.648$, which identifies the occurrence of percolation in
the network, way before the SN transition has happened, and
thus it may be used as a signal that a dynamical transition is
coming. The distance in the parameter space between that signal
and the transition depends on which threshold is used to build
the network, but there is an optimal value (see below).


\begin{figure}
\begin{center}
\includegraphics[width=0.7\columnwidth]{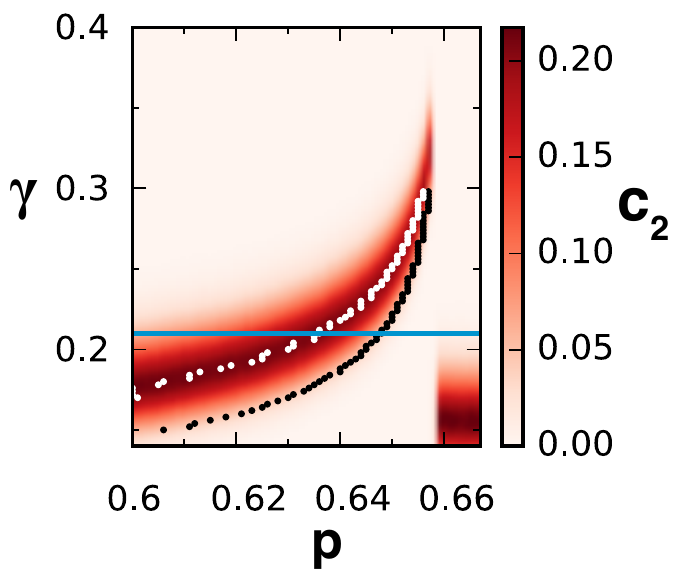}
\caption{Role of the correlation threshold $\mathbf{\gamma}$.
$c_2$ values, as given by the color bar, as a function of the
control parameter $p$ and the threshold $\gamma$ used to build the
functional network for the LE model. The white dots give
the locations of the $c_2$ maxima, while the black dots mark the maxima of
$\langle s^2 \rangle$ (percolation transition). The dynamical
sudden jump occurs at $p_d=0.658$. The horizontal
blue line identifies the value of $\gamma = 0.21$ used in Fig. \ref{figLE}. }
\label{figLEgamma}
\end{center}
\end{figure}

The probabilities $c_2$, $c_5$ and $c_9$ for the networks built
from the LE correlations are shown in Fig. \ref{figLE}d. Their
maxima clearly anticipate the percolation transition signaled
by $S_1$ and $\langle s^2 \rangle$, which gives itself an early
warning of the SN bifurcation. The peak in $c_2$ appears at
$p_{2}=0.635$, anticipating $p_d$ more than twice as early as
the percolation transition.

Building functional networks involves to fix the correlation
threshold $\gamma$ above which two elements are considered as
linked. Figure \ref{figLEgamma} shows how the percolation
transition and the values of $c_2$ depend on both $p$ and
$\gamma$. If $\gamma$ is very high, the network is never
connected and there is no signal. Similarly, if $\gamma$ is
very low the network is always fully connected and there is no
hint of the bifurcation. However, as shown in the figure, there
is a range of values of $\gamma$ where the percolation
transition and its associated early warning signals appear. For
a fixed value of $\gamma$ the peak of $c_2$ (white dots) always
occurs earlier than the percolation transition (black dots). As
$\gamma$ decreases, the peak of $c_2$ occurs at earlier values
of $p$. The curve of $c_2$, however, widens and the resolution
in the location of the peak gets poorer. The lowest (optimal)
value of $\gamma$ at which the peak can be distinguished marks
thus the earliest warning signal that can be obtained for the
bifurcation. Note that this does not imply that the method only works for a fixed value of $\gamma$. In an empirical situation, one may need to explore this parameter, but there is a range of values of $\gamma$ over the optimal that will provide valid early warning signals for the transition.

\subsection*{The Ginzburg-Landau equation}


\begin{figure}
\begin{center}
\includegraphics[width=0.7\columnwidth,clip=true]{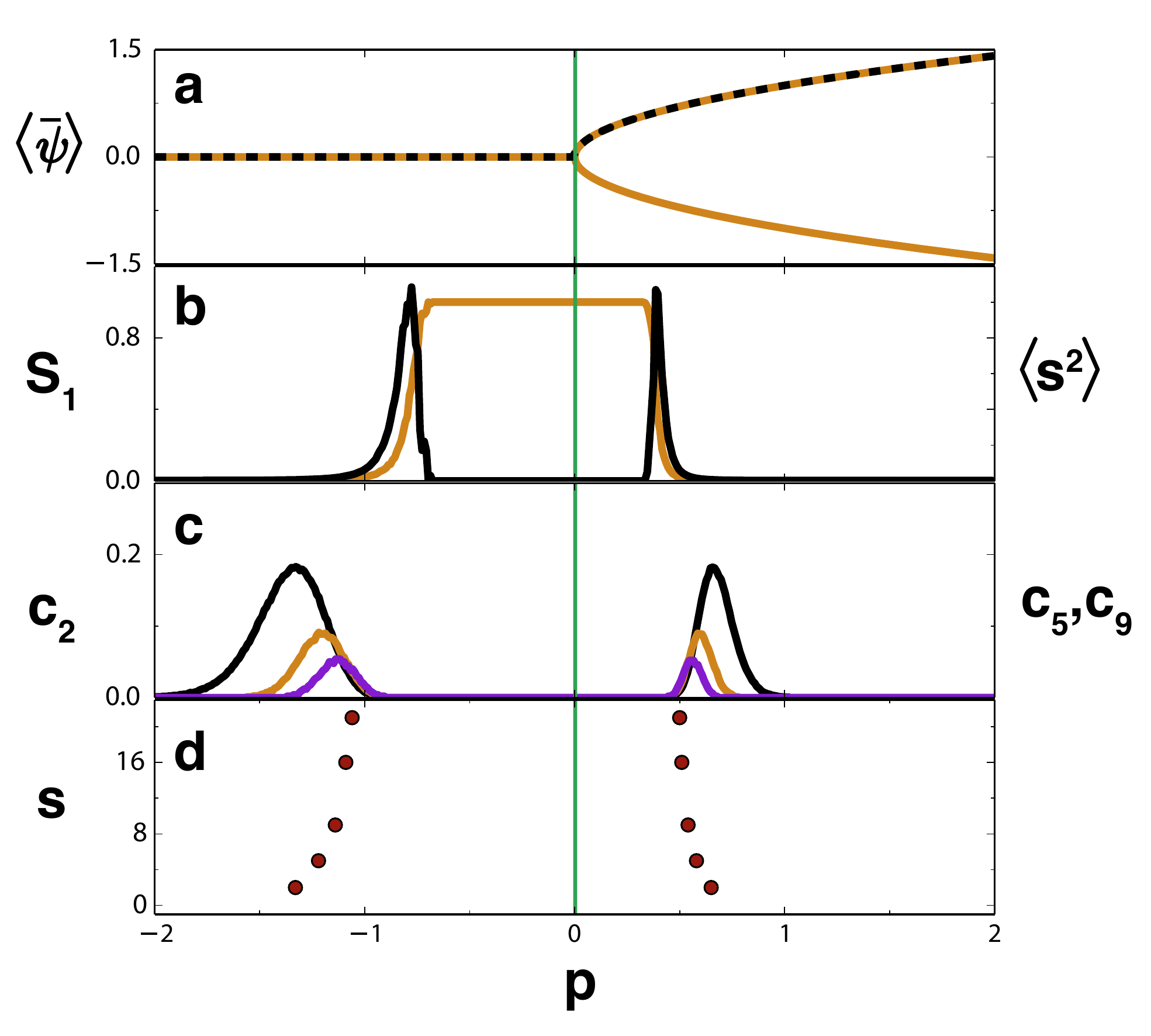}
\caption{The GL model has a continuous transition at the
critical value $p_d = 0$ marked by the green line. a) The theoretical (yellow)
and the numerically obtained (black dash line, further averaged over $R=1000$ temporal
snapshots) homogeneous value of the field $\psi$.
b) $S_1$ (orange) and $\langle s^2\rangle$ (black).
c) The quantities $c_2$ (black), $c_5$ (orange), and $c_9$ (purple).
d) Circles indicate the values of $p$, $p_s$ with $s=2,5,9,16$ and $21$,
for which $c_s$ attains its respective maximum.
Functional networks were built using a threshold of $\gamma = 0.25$.}
\label{figGL}
\end{center}
\end{figure}

To prove the generality of the precursors, we analyze a
different system, the time-dependent Ginzburg-Landau (GL)
equation or model A \cite{Hohenberg1977} describing, for
example, transitions in anisotropic ferromagnets. It is a
paradigmatic model experiencing a continuous transition, namely
a supercritical pitchfork bifurcation. We study the
one-dimensional version for the magnetization $\psi(x,t)$:
\begin{equation}
\frac{\partial\psi(x,t)}{\partial t} = p\psi(x,t) - \psi(x,t)^3 + \epsilon\nabla^2\psi(x,t)  + \eta(x,t).
\label{Gi}
\end{equation}
As before, $\epsilon$ is diffusive coupling and $\eta$ is an
additive uncorrelated noise uniform in $[-a,a]$. We discretize
equation (\ref{Gi}) into $N=5000$ nodes, and take
$\epsilon=1.5$, $a=0.01$. The integration parameters $dt$ and
$dx$ are as for the LE model. A continuous transition from zero
magnetization occurs when increasing the control parameter $p$
above $p_d=0$. It is seen in Fig. (\ref{figGL}a), which shows
temporal averages of the spatial mean field $\bar\psi$.

We compute spatial correlations and build functional networks
from $R=1000$ snapshots using a threshold $\gamma=0.25$. The
corresponding percolation quantifiers are plotted in Fig.
\ref{figGL}. Here correlations continuously build-up when
increasing $p$ towards the critical point at $p_d=0$ and,
unlike the previous discontinuous transition, continuously
decrease after crossing it. Therefore, there are now two
percolation transitions in the functional networks: one at each
side of $p_d$, as seen by the indicators $S_1$ and $\langle s^2
\rangle$ in Fig. \ref{figGL}b. Figure \ref{figGL}c depicts the
probabilities $c_2$, $c_5$ and $c_9$, with maxima giving a
clear warning further away from $p_d$. Figure \ref{figGL}d
displays the values $p_s$ corresponding to these maxima in
$c_{s}$. There are two successions of peaks converging to the
percolation transitions occurring before and after the
bifurcation.

\begin{figure}
\begin{center}
\includegraphics[width=0.7\columnwidth]{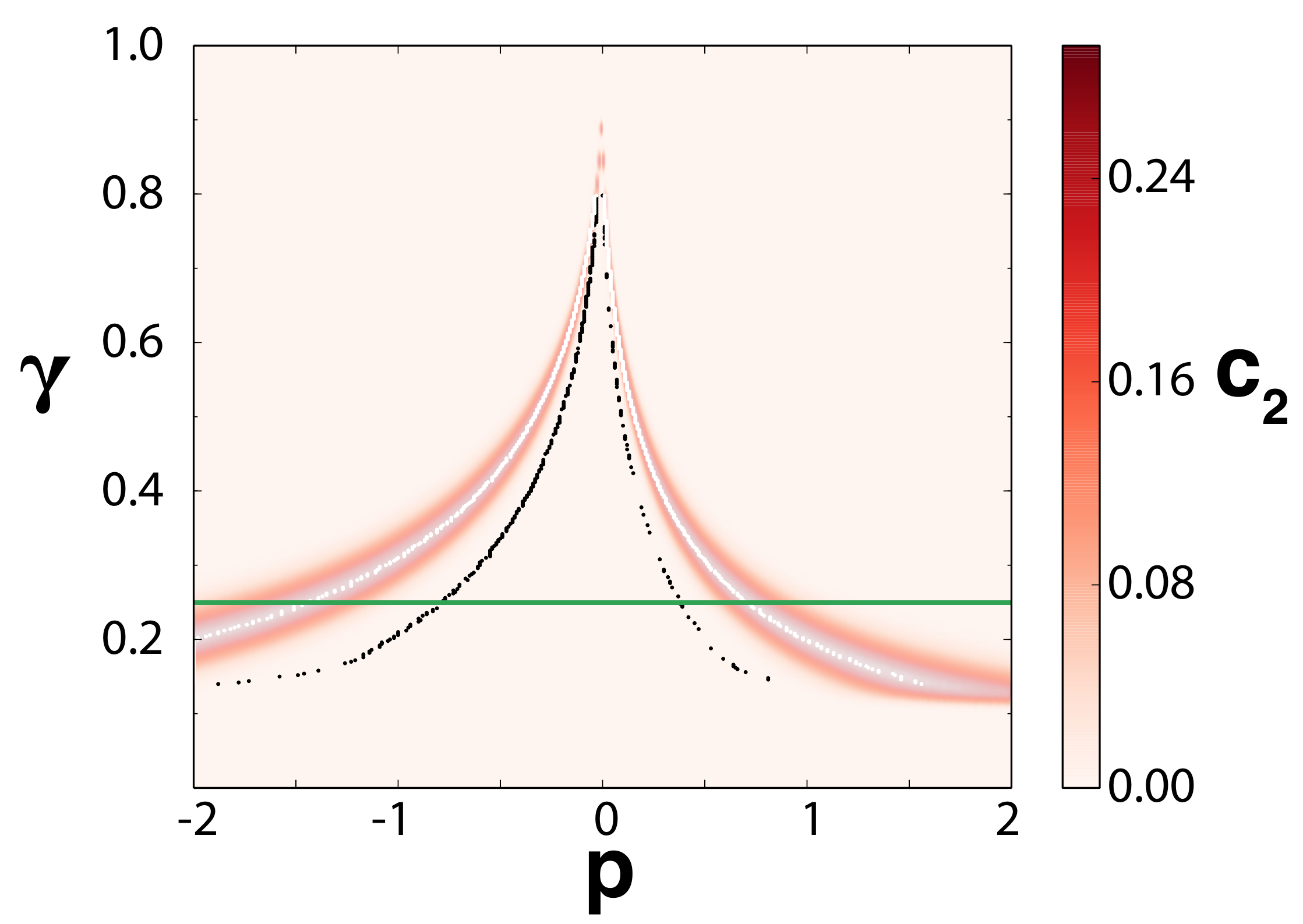}
\caption{The $c_2$ values, as given by the color bar as a
function of the control parameter $p$ and the threshold
$\gamma$ used to build the functional network for the
Ginzburg-Landau system. The continuous pitchfork bifurcation
occurs at $p_d=0$. Black dots indicate the maxima of $\langle
s^2 \rangle$, which locate the percolation transition. White
dots locate the maxima of $c_2$. We see how the percolation
transition and its precursor $c_2$ anticipate in different
amounts (when increasing $p$ from the low $p$ state, or when
decreasing $p$ from the high $p$ state) the dynamical
bifurcation. The horizontal green line identifies the value
$\gamma=0.25$ for which Fig. 3 of the main text was
constructed.} \label{figGLgamma}
\end{center}
\end{figure}

Figure \ref{figGLgamma} shows what happens to the percolation
transition when building the networks for different thresholds
$\gamma$. The figure displays the maxima of $\langle s^2
\rangle$ (which locate the percolation transitions) and the
values of $c_2$, in the ($\gamma, p$) parameter space. We see
that, when increasing $p$, the maximum in the precursor $c_2$
anticipates the percolation transition, which itself
anticipates the pitchfork bifurcation at $p=0$. At the other
side of the transition, when decreasing $p$ from the high $p$
state, the maximum in $c_2$ also occurs before the percolation
transition, which also anticipates the dynamical transition. As
in the LE case the amount of anticipation is larger for lower
$\gamma$, until the signal disappears.

\subsection*{The Lorenz'96 system}

\begin{figure}
\begin{center}
\includegraphics[width=0.7\columnwidth,clip=true]{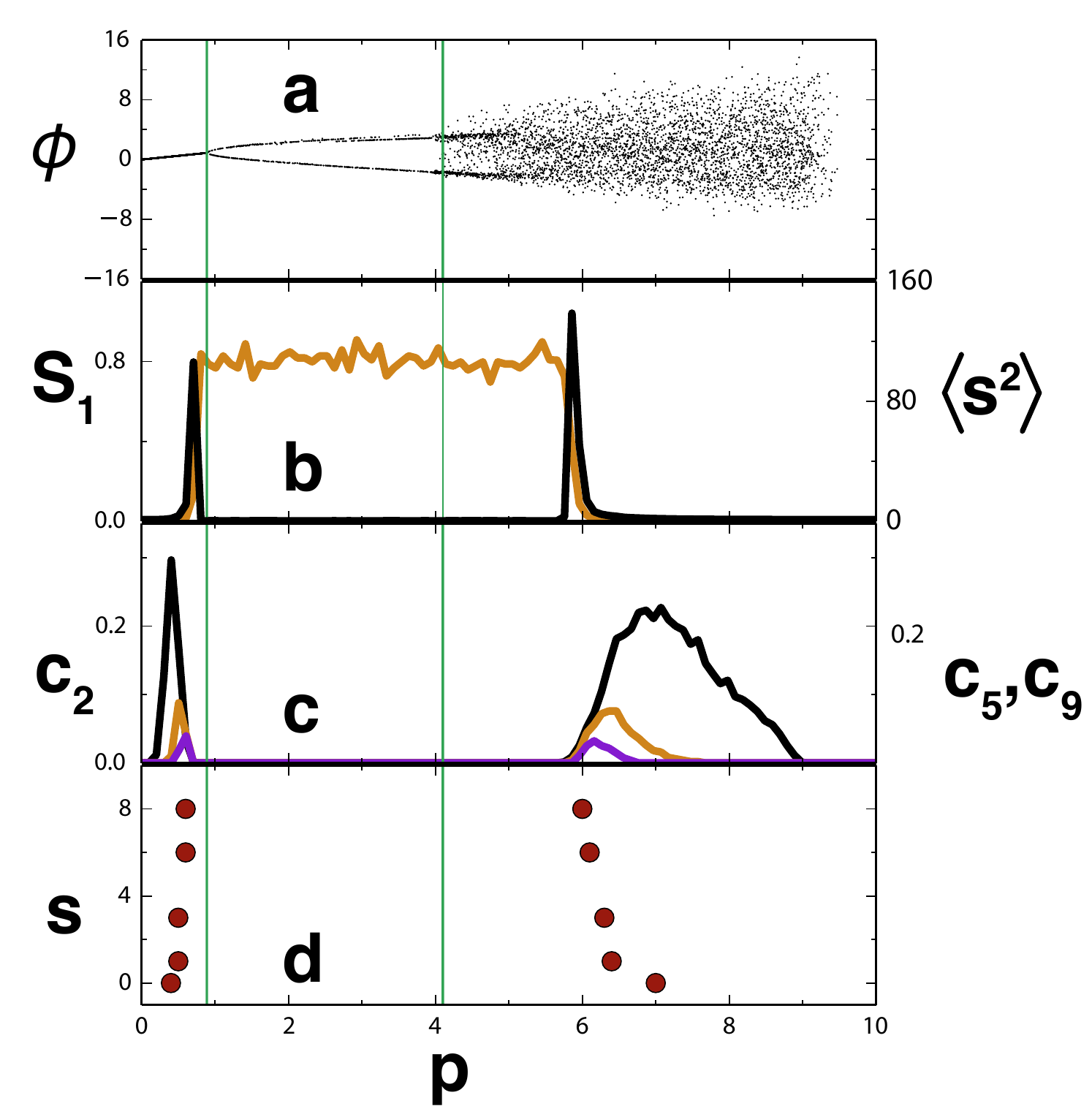}
\caption{Panel a) shows the bifurcation diagram of
the Lorenz'96 model, equation (\ref{Lo}), constructed from a
2-oscillator Poincar\'{e} section (see text). The transition to
traveling waves at $p_{1} = 8/9$ and to spatiotemporal chaos
($p_{2}\approx4.1$) are shown as vertical green lines.
Functional networks were built using a threshold of $\gamma =
0.16$. The percolation indicators $S_1$ (orange) and
$\left<s^2\right>$ (black) are displayed in panel b).
Panel c) shows $c_2$ (black), $c_5$ (orange), and $c_9$
(purple). In d), the circles indicate the value of $p$,
$p_s$ with $s=2,5,9,16$ and $21$, for which the $c_s$ curves
attain their respective maxima. These indicators reveal a phase
of percolated correlations in a parameter region which includes
the interval $[p_{1},p_{2}]$, flanked by two percolation
transitions. The curves have been further averaged over 100
realizations of the random noise and initial condition.}
\label{figL96}
\end{center}
\end{figure}

\begin{figure}
\begin{center}
\includegraphics[width=0.7\columnwidth]{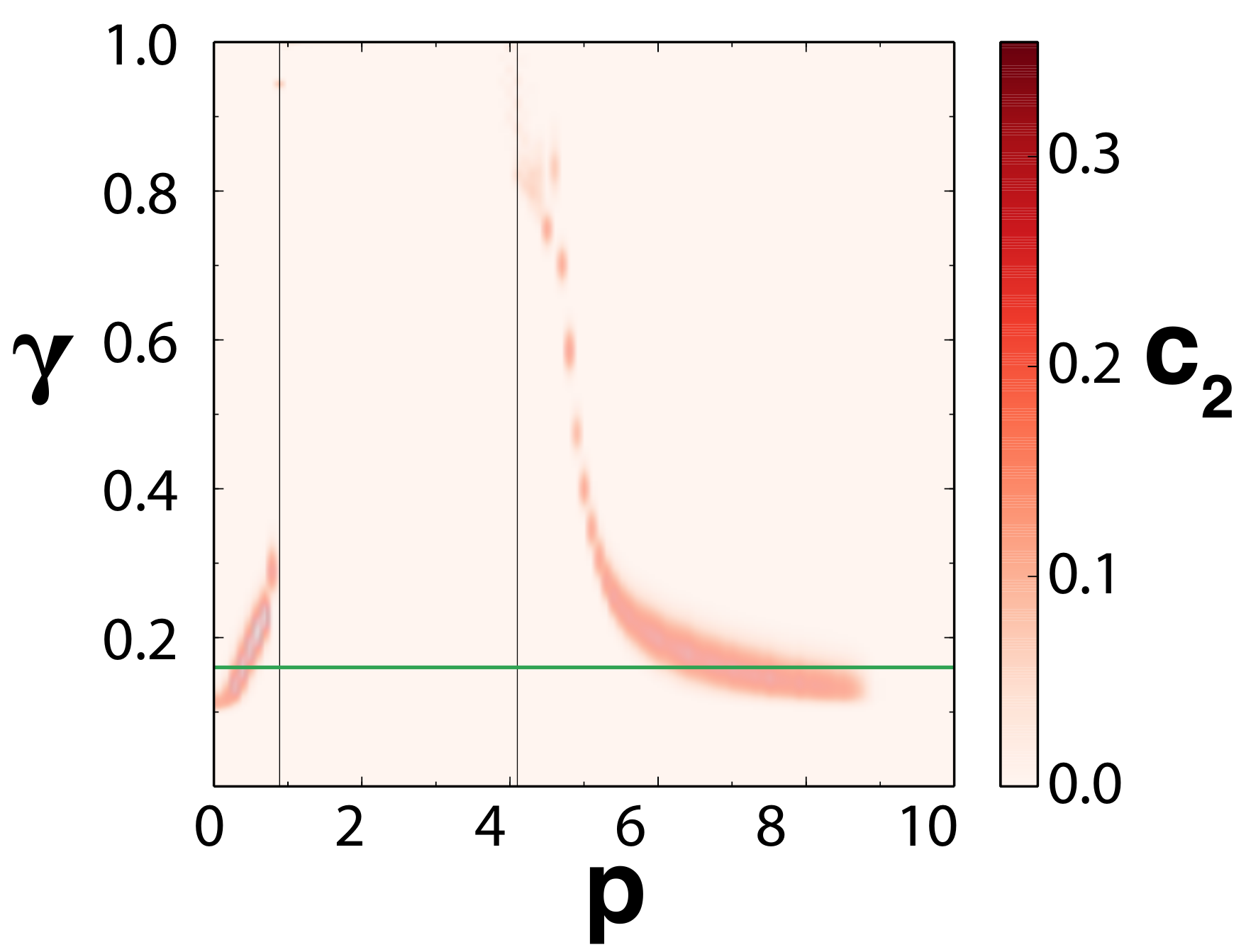}
\caption{The $c_2$ values, as given by the color bar, as a
function of the control parameter $p$ and the threshold
$\gamma$ used to build the functional network for the Lorenz'96
system. The dynamical transitions to traveling waves and to
chaos are indicated by the vertical black lines. The horizontal
green line identifies the value $\gamma=0.16$ for which Fig.
\ref{figL96} was constructed.} \label{figL96gamma}
\end{center}
\end{figure}

Coupled chaotic oscillators display a large variety of
dynamical regimes. Thus, due to the different bifurcations
present in those models, they are an excellent test bed to
prove the generality of the network-based percolation
precursors methodology. Here we consider the Lorenz'96 model
\cite{Lorenz1996}. It was proposed by E. Lorenz as a simplified
framework to investigate atmospheric predictability. It reads:
\begin{equation}
\frac{d\psi_k(t)}{dt}=\left[\psi_{k+1}(t) - \psi_{k-2}(t)\right]\psi_{k-1}(t) - \psi_k(t) + \eta_k(t) + p \ .
\label{Lo}
\end{equation}
$\psi_k(t)$ is meant to represent the values of some
atmospheric variable at different locations $k$, $k=1\ldots N$,
arranged in a one dimensional ring around the globe (an thus
having periodic boundary conditions). The structure of equation
(\ref{Lo}) contains some of the main elements of fluid
dynamics, namely dissipation, external forcing, and quadratic
non-linearity through an advection-like term. In addition to
the constant forcing $p$, which will be our control parameter,
we include an additive stochastic perturbation $\eta_k(t)$
uncorrelated in space and time. It is implemented here by
adding independent random numbers uniform in $[-a,a]$ ($a=0.1$)
after each time step ($dt=1/64$) of a fourth-order Runge-Kutta
method which is used to integrate the rest of the terms. We
focus in the behavior for $N=2500$ elements or oscillators. In
the absence of the random forcing, three dynamical regimes are
easily identified \cite{Karimi2010}: i) For small $p$,
$p<p_1=8/9$, the system stabilizes in the homogeneous fixed
point $\psi_{k}(t) = p , \forall \ k, \ t$. ii) A Hopf
bifurcation occurs at $p=p_1$ so that for intermediate values
of $p$, $8/9<p<4.1$ the system is in a traveling-wave state.
iii) Beyond $p>p_2\approx 4.1$ the system becomes
spatiotemporally chaotic.

To display the bifurcations observed when integrating equation
(\ref{Lo}) for $N=2500$ elements we have calculated a
Poincare's transversal section in the subspace of two
contiguous oscillators: one of the oscillators, say $\psi_1$,
is monitored and when it crosses the value $\psi_1=1$ in the
increasing direction the value of the contiguous oscillator,
say $\phi=\psi_2$ is recorded and displayed. The bifurcation
diagram showing the values of these sections $\phi$ is plotted
in Fig. \ref{figL96}a as a function of $p$. The three regimes
described above for the deterministic system are readily
identified here also.

Functional networks were constructed from $R=1000$ snapshots by
interpreting the locations $k$ as nodes and assigning links
between pairs of nodes when the Pearson correlation is larger
than a threshold $\gamma=0.16$. Figure \ref{figL96}b shows the
quantities $S_1$, the size of the largest cluster, and
$\left<s^2\right>$, the mean cluster size excluding the largest
one, for such network. Panels c) and d) display the properties
of several $c_s$, the probabilities of randomly chosen nodes to
pertain to clusters of size $s$. These figures clearly identify
the presence of a percolated phase at intermediate values of
$p$, started and ended by two percolation transitions. Figure
\ref{figL96gamma} shows the quantity $c_2$ in the ($\gamma, p$)
parameter plane for this model. We see that for increasing
threshold the maxima of $c_2$ approach the locations $p_1$ and
$p_2$ at which traveling waves are born via a Hopf bifurcation
and at which they destabilize into chaotic behavior,
respectively. Thus, the percolating phase is a manifestation of
the long-range coherence of the traveling wave state, whereas
correlation length remains small in the homogeneous and in the
chaotic regime. The quantity $c_2$ (and indeed the other $c_s$)
clearly anticipates the first bifurcation when increasing $p$.
It also largely anticipates the occurrence of a chaos-order
transition when decreasing $p$ from large values.

\subsection*{Percolation in sea temperature networks during El Ni\~{n}o events}

\begin{figure}
\begin{center}
\includegraphics[width=.95\columnwidth]{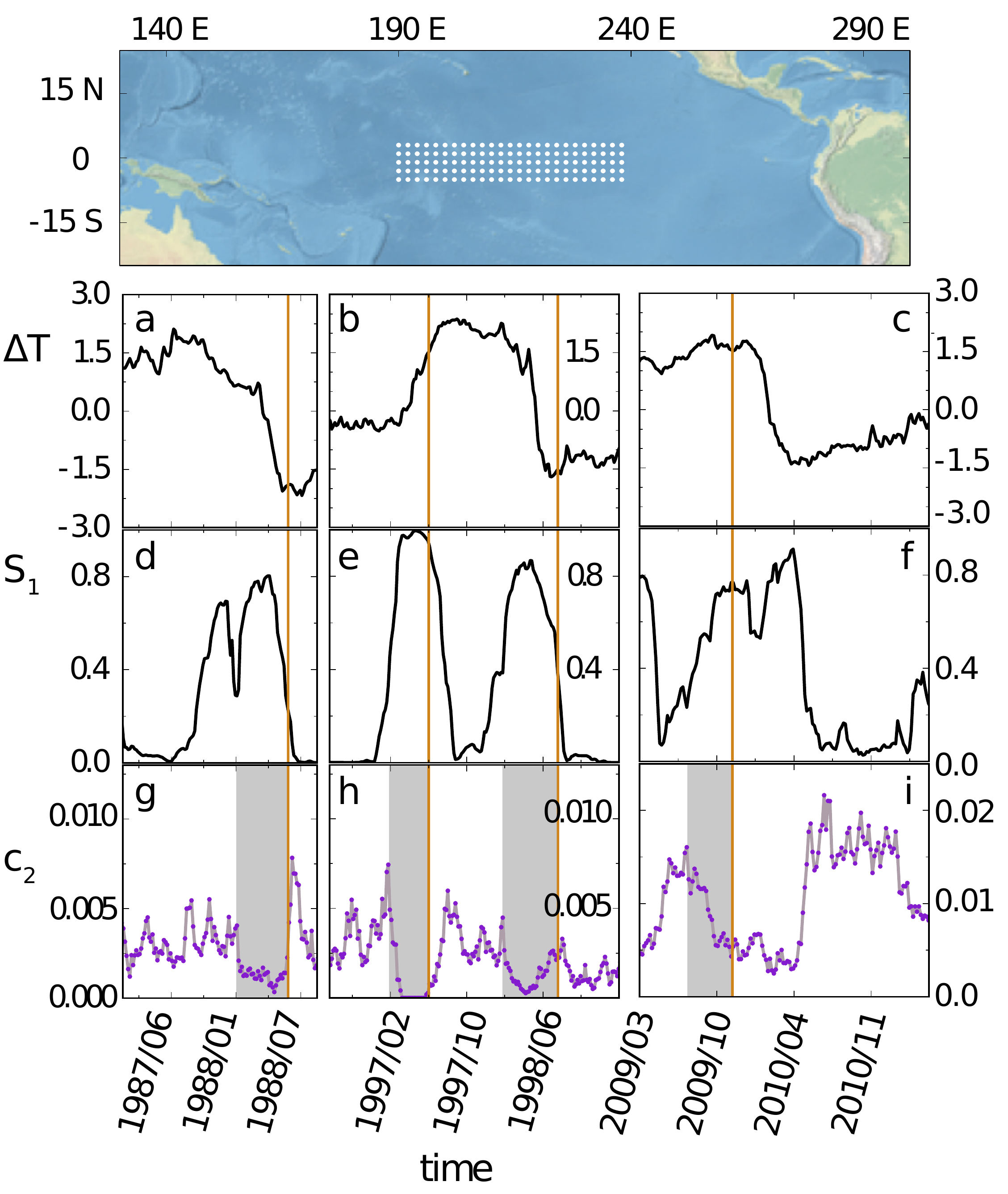}
\caption{Application to El Ni\~no phenomenon. The upper panel shows a map
of the area over which the mean Sea Surface Temperature $T$ is
monitored in the NINO3.4. Points denote locations used here as nodes
in a functional network. The map was generated with Cartopy 0.11.0  \cite{cartopy}. Four events, two La Ni\~{n}a (cold) and two
El Ni\~{n}o (warm), are shown in the time axis of panels
a) to i). Conventional starting dates of the events are marked
by vertical orange lines. In a), b) and c), the sea surface temperature $T$ is shown as a
function of time. A functional network is constructed from correlations
at $\gamma= 0.99992$ for $1987-1989$ and $1996-1998$, and  $\gamma = 0.9986$ for $2009$. The size of the giant component $S_1$ is shown in panels d), e) and f),
showing percolating phases at a plateau, flanked by two percolation transitions,
which occurs before each of the events. Panels g), h) and i) show
$c_2$ in the same time frame. Peaks in $c_2$
flank both sides of the percolation plateaux, in a manner similar to the
Ginzburg-Landau case shown in Fig. \ref{figGL}. The time by which the peaks of $c_2$ anticipate the conventional starting date of the event is marked in gray.}
\label{figENSO}
\end{center}
\end{figure}

To test the behavior of our precursors in observed real
situations, we analyze sea surface temperature data from the
region of the Pacific used to compute the NINO3.4 index
\cite{Ludescher2013}. El Ni\~{n}o-Southern Oscillation
\cite{Sarachick2010,Dijkstra2006} is the dominant variability
mode in present-day climate, characterized by rather irregular
(with average period of about 4 years) warm (El Ni\~{n}o) and
cold (La Ni\~{n}a) episodes departing from the long-term mean
temperature in the equatorial Pacific. These oscillations are
related to the presence of a Hopf bifurcation in the coupled
atmosphere-ocean system \cite{Sarachick2010,Dijkstra2006}. The
bifurcation can be crossed or just approached, being then the
oscillation excited by noise. In both cases there should be a
build-up of correlations that would become visible in
functional networks constructed from temperature time series.
In this case there is no control parameter to fix, but rather
the equatorial Pacific evolves in time, coupled to the seasonal
cycle, leading to changing spatial correlations. We will see
that, despite this lack of control, and without using any
information on the underlying dynamics, our approach is able to
find precursors of the relevant El Ni\~{n}o-La Ni\~{n}a events.

Sea Surface Temperatures were obtained from the ERA-interim
reanalysis of the European Centre for Medium-Range Weather
Forecasts\cite{ECSMs}, with daily temporal resolution and a
spatial resolution of $\Delta x = 0.125^\circ$, in the range of
years $1979-2014$ (Fig. \ref{figENSO}). Daily functional
networks at day $t$ were built from these time series computing
the Pearson correlation with a time window of $R=200$ days (100
days before and 100 days after time $t$). The quantities
plotted in Fig. \ref{figENSO} are further averaged over 5 days.

In the Figure \ref{figENSO}, we have focused on three different periods:  $1987-1989$, during
which a strong La Ni\~{n}a occurred, $1996-1998$, featuring
one El Ni\~{n}o-La Ni\~{n}a pair, and a recent El Ni\~no in 2009.
In contrast to the previous examples, this system is empirical and the contribution of the noise is more difficult to assess than in a model equation. Therefore, the systematic search for the optimal $\gamma$ has not been  performed. Nevertheless, the space of values of $\gamma$ and how they affect $c_2$ have been explored in Supplementary Figure S1. Interestingly, it seems that the range of values of $\gamma$ necessary to observe peaks in $c_2$ have moved toward lower $\gamma$ in the early $2000$s. We have fixed two values of the threshold to produce Figure \ref{figENSO}: $\gamma = 0.99992$ for the events of $1987-1989$ and $1996-1998$, and  $\gamma = 0.9986$ for the event of $2009$. In the two cases, this is where a nice compromise between signal-and-noise is found. In a practical situation, the selection of $\gamma$ is not a post-hoc process: one can have a clear idea of the range of values to use from the previous events. Once $\gamma$ is set at a fixed value in this range, if $c_2$ shows a peak, followed by a sequence of peaks of $c_3$, $c_5$, etc., the system is very likely going towards a new El Ni\~no/Ni\~na event. The panels a, b and c of Fig. \ref{figENSO} display the variations of the ocean superficial temperature and also the moments at which an El Ni\~no or La Ni\~na event are officially declared are marked with a vertical orange line. The panels d, e and f, on the other hand, depict the time evolution of the size of the largest connected component $S_1$, which peaks before or on the arrival of the event. $c_2$, represented in the lower panels (g, h and i), also shows maxima way before the corresponding peak of $S_1$. The anticipatory period since the $c_2$ peaks to El Ni\~no (La Ni\~na) event is marked in gray in every plot. It corresponds to 240 days in $1988$ (Fig.\ref{figENSO}g), 125 days in $1997$ (Fig.\ref{figENSO}h), 175 days in $1998$ (Fig.\ref{figENSO}h) and 115 days in $2009$ (Fig.\ref{figENSO}i).

\section*{Conclusions}

In summary, we have shown that consideration of the percolation transition
in functional networks constructed from spatial correlations in
extended systems provides powerful anticipatory tools for their
dynamical regime shifts. Precursors of the percolation
transition itself, such as the probabilities $c_s$ for random
nodes to belong to small clusters of size $s$, add extra anticipatory
range.  This is done by introducing a mesoscopic view of the system, instead of using global or local perspectives as the ones used in previous methods using the system dynamics slowing down, the spatial variability of the order parameter or the clustering and degree distribution of the functional network. Furthermore, the sequence of peaks of $c_s$ provide extra information on the distance still remaining to the (percolation) transition. We note that $\gamma$ and $s$ are methodological
parameters, so that they can be explored even when far from the
dynamical transition. Despite of the fact that there exists an optimal $\gamma$ for which the anticipatory power is largest, there is typically a wide range of values of $\gamma$ for which the method works. Furthermore, the sequence of peaks of $c_s$ can give a good hint on the distance to
the approaching transition.

The tools presented here work in a variety of transition types,
the condition being the increase of spatial correlations when
approaching the transition point. This happens generally at
least for systems close to bifurcations characterized by
critical slowing down and with spatial locations coupled by
diffusion. Most of the local bifurcations types satisfy the
critical slowing down criterion (see discussion in \cite{Thompson2011,Dakos2015}). Although
diffusive coupling is sufficient to provide increasing spatial
correlations when combined with critical slowing down, it is by
no means necessary, as the example of the Lorenz'96 model (for
which spatial coupling is of advection type rather than
diffusive) shows.

Through this paper we have focused in spatially embedded
complex systems, in which network nodes are associated to
spatial locations. Since the only information needed to apply
our framework is a set of time series coming from different
network nodes, we expect our approach to be also useful in more
general network systems experiencing regime transitions
\cite{Dorogovtsev2008}, beyond the spatial ones. In addition we
have shown that the percolation-based precursors can be used
even in cases, such as El Ni\~{n}o events, where very little
or no information on the underlying dynamics is available. This is, therefore, a fresh perspective on a known phenomena with the bonus of offering a method that  can become instrumental in the monitoring and
management of complex systems.

\section*{Acknowledgements}
V.R.-M. was supported by the European Commission Marie-Curie ITN
program (FP7-320 PEOPLE-2011-ITN) through the LINC project
(Grant no. 289447). We also acknowledge support from FEDER and
Spanish Ministry of Economy and Competitiveness (MINECO) through the project INTENSE@COSYP
(FIS2012-30634) and from the European Commission through
project LASAGNE (FP7-ICT- 318132). J.J.R. acknowledges funding from the Ram\'on y Cajal program of MINECO.

\section*{Author contributions statement}

All the authors conceived the method, V.R.-M. conducted the numerical simulations and data treatment, all the authors analysed the results, contributed in the writing and reviewed the manuscript.

\section*{Additional information}

The authors declare no competing financial interests.

\newpage

\renewcommand{\thefigure}{S\arabic{figure}}
\setcounter{figure}{0}  

\begin{center}

\ \vspace{5cm}

\Huge{SUPPLEMENTARY INFORMATION FOR PERCOLATION-BASED
PRECURSORS OF TRANSITIONS IN EXTENDED SYSTEMS}

\ \vspace{2cm}

\Huge{by V\'{\i}ctor Rodr\'{\i}guez-M\'{e}ndez, V\'{\i}ctor M.
Egu\'{\i}luz, Emilio Hern\'{a}ndez-Garc\'{\i}a and Jos\'e J.
Ramasco}

\end{center}

\begin{figure}[b]
  \centering
  \includegraphics[width=15cm]{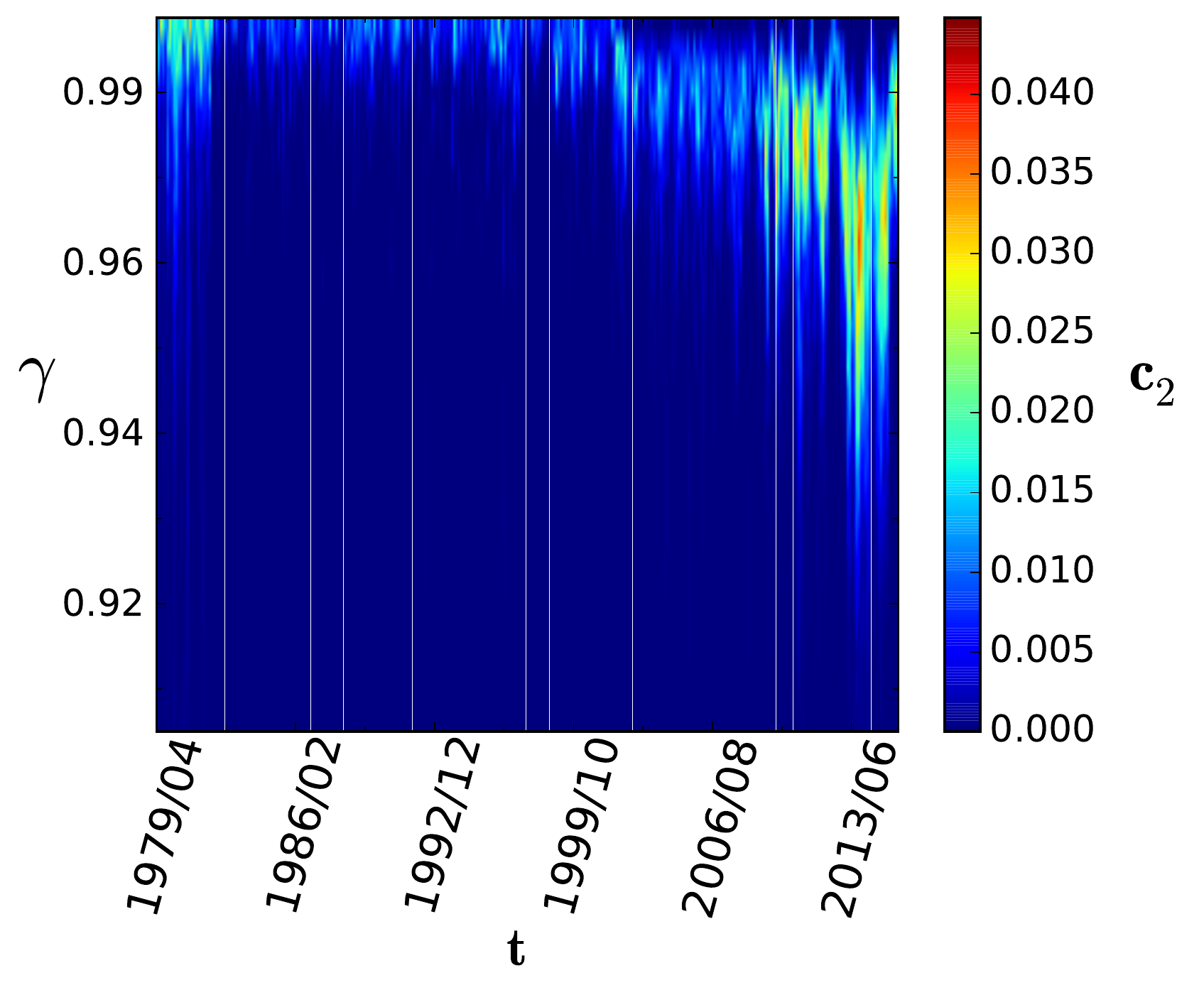}
  \caption{\label{fig:subcat}  Heatmap with an exploration of the fraction of nodes in
  clusters of size 2, $c_2$, as a function of time and of $\gamma$ for functional networks
  extracted daily over $30$ years of data on the mean Sea Surface Temperature $T$ monitored in the NINO3.4. The spatial resolution is $0.5$ arc degrees.}
\end{figure}

\end{document}